\documentclass[twocolumn,english,prl,amsmath,amssymb,preprintnumbers]{revtex4}
\usepackage[T1]{fontenc}
\usepackage[latin9]{inputenc}
\usepackage{color}
\usepackage{bm}
\usepackage{amstext}
\usepackage{esint}
\usepackage{soul}% \st{text} provides strikeout of "text"
\usepackage{graphicx}
\usepackage{dcolumn}
\usepackage{bigstrut}
\usepackage{rotating}    

\makeatletter
\@ifundefined{textcolor}{}
{%
 \definecolor{BLACK}{gray}{0}
 \definecolor{WHITE}{gray}{1}
 \definecolor{RED}{rgb}{1,0,0}
 \definecolor{GREEN}{rgb}{0,1,0}
 \definecolor{BLUE}{rgb}{0,0,1}
 \definecolor{CYAN}{cmyk}{1,0,0,0}
 \definecolor{MAGENTA}{cmyk}{0,1,0,0}
 \definecolor{YELLOW}{cmyk}{0,0,1,0}
}

\def\be{\begin{equation}}
\def\ee{\end{equation}}
\def\bea{\begin{eqnarray}}
\def\eea{\end{eqnarray}}
\def\bse{\begin{subequations}}
\def\ese{\end{subequations}}

\usepackage{dcolumn}% Align table columns on decimal point
\usepackage{bm}% bold math
\usepackage{amsfonts}
\draft

\makeatother

\usepackage{babel}
\begin{document}
\preprint{Phys. Rev. Lett. {\bf 115}, 020402 (2015)}
%PRL LD15099

\title{Third Law of Thermodynamics and the Shape of the Phase Diagram for Systems with a First-Order Quantum Phase Transition}

\author{T. R. Kirkpatrick$^{1}$ and D. Belitz$^{2,3}$}

\affiliation{$^{1}$ Institute for Physical Science and Technology,and Department
of Physics, University of Maryland, College Park, MD 20742, USA\\
$^{2}$ Department of Physics and Institute of Theoretical Science,
University of Oregon, Eugene, OR 97403, USA\\
$^{3}$ Materials Science Institute, University of Oregon, Eugene,
OR 97403, USA\\
  }

\date{\today}
\begin{abstract}
The third law of thermodynamics constrains the phase diagram of systems with
a first-order quantum phase transition. For zero conjugate field, the 
coexistence curve has an infinite slope at $T=0$. If a tricritical 
point exists at $T>0$, then the associated tricritical wings are perpendicular to the 
$T=0$ plane, but not to the zero-field plane. These results are based on the third law and 
basic thermodynamics only, and are completely general. As an explicit example we
consider the ferromagnetic quantum phase transition in clean metals, where a first-order quantum
phase transition is commonly observed.
\end{abstract}
% 599 characters
%
\maketitle

First-order phase transitions are ubiquitous in nature, the solid-to-liquid and liquid-to-gas transitions being
the most commonly observed ones. Another common example of a first-order transition is the ferromagnetic
transition below the Curie temperature as a function of an external magnetic field. First-order transitions
are characterized by a coexistence curve in the phase diagram along which the two phases coexist in
thermodynamic equilibrium. (The coexistence curve may be the projection of a higher-dimensional
coexistence manifold into a particular plane in the phase diagram.) 

It has long been known that the curvature of the coexistence curve is determined by the discontinuities
of certain observables across it. The Clapeyron-Clausius (CC) equation relates the slope of the coexistence curve
in the pressure-temperature ($p\,$-$T$) plane to the discontinuities of the entropy and the volume \cite{Landau_Lifshitz_V_1980}:
\be
\left(\frac{dp}{dT}\right)_H = \frac{\Delta s}{\Delta v}\quad,
\label{eq:1}
\ee
where $\Delta s = s_1 - s_2$ and $\Delta v = v_1 - v_2$ with $s_{1,2}$ and $v_{1,2}$ the specific entropy and
volume per particle, respectively, in the two phases. For definiteness, let 1 and 2 label the ordered and disordered
phases, respectively, and for later reference we indicate that {an appropriate external} field $H$, if any, is held constant 
in taking the derivative.

\begin{figure}[t]
\includegraphics[width=8cm]{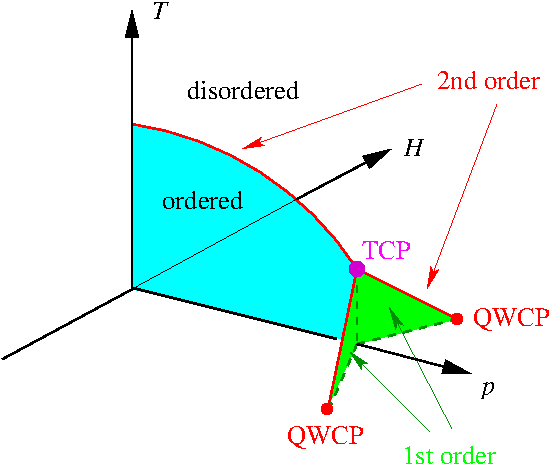}
\caption{{Schematic phase diagram showing a line of first-order transitions at low $T$ separated from a line of second-order
              transitions at higher $T$ by a tricritical point (TCP). In a nonzero conjugate field $H$ tricritical wings emerge from the TCP.} 
              These are surfaces of first-order transitions that are bounded by lines of second-order transitions
              and terminate in two quantum wing-critical points (QWCP) in the $T=0$ plane.}
\label{fig:1}
\end{figure}
The CC equation (\ref{eq:1}) and its analogs in different planes of the phase diagram are very general,
as they rely only on basic thermodynamic arguments. In this Letter we show that for {\em quantum} phase
transitions, when combined with the third law of thermodynamics, they provide interesting constraints on
the shape of the phase diagram. {We will consider a pressure-driven transition at $T=0$ that is first order,
remains first order at low $T$, and turns second order at higher $T$ via a tricritical point (TCP).} The schematic phase
diagram in the space spanned by $T$, $p$, and $H$, {where $H$ is the field conjugate to the order parameter,}
is shown in Fig.~\ref{fig:1}. {As we will see, the detailed shape of this phase diagram at low $T$ is constrained by
thermodynamics. Our arguments leading to this conclusion are completely general; however, as an explicit example 
we will discuss the} quantum ferromagnetic transition in clean metals {\cite{Brando_et_al_2015}. Another example
of a first-order quantum phase transition with a TCP in the phase diagram is the Ising antiferromagnet dysprosium
aluminum garnet \cite{Giordano_Wolf_1977}.}

We are interested
in a system with $T$, $p$, and $H$ as independent variables. Denoting the order parameter by $M$, the
appropriate thermodynamic potential is the generalized Gibbs free energy \cite{Gibbs_free_energy_footnote}
\bse
\label{eqs:2}
\bea
\tilde{G} &=& U - TS + pV - HM
\nonumber\\
   &=& \mu N\ ,
\label{eq:2a}
\eea
whose differential is
\be
d\tilde{G} = -SdT + Vdp - MdH + \mu\, dN\ .
\label{eq:2b}
\ee
\ese
Here $S$, $V$, and $\mu$ are the system's entropy, volume, and chemical potential, respectively,
and $N$ is the particle number. From Eqs.~(\ref{eqs:2}) we obtain the Gibbs-Duhem relation
\be
d\mu = d\tilde{g} = -sdT + vdp - mdH\ ,
\label{eq:3}
\ee
where $\tilde{g}$, $s$, $v$, and $m$ are the generalized Gibbs free energy, entropy, volume, and {order parameter} per particle, 
respectively. On the coexistence 
curve the chemical potentials of the two phases must coincide. Using this condition with Eq.~(\ref{eq:3}) at fixed 
{external} field leads to Eq.~(\ref{eq:1}).  An analogous argument yields
\be
\left(\frac{dT}{dH}\right)_p = - \frac{\Delta m}{\Delta s}\ .
\label{eq:4}
\ee

The CC equations (\ref{eq:1},\ref{eq:4}) are completely general. When applied to a quantum
phase transition, the third law provides the additional constraint $\Delta s(T\to 0) \to 0$. To be
specific, let us assume that {in either phase the entropy vanishes
as $s(T\to 0) = \gamma\,T^n$. In particular, if the phases are Fermi liquids (see below) then $n=1$ and $\gamma$
is the specific-heat coefficient.}
%Obvious generalizations apply to cases where one or both phases are not Fermi liquids.
For asymptotically low temperatures we thus have
\bse
\label{eqs:5}
\bea
\left(\frac{dT}{dp}\right)_H &=& \frac{1}{{T^n}}\,\frac{\Delta v}{\Delta\gamma}\ ,
\label{eq:5a}\\
\left(\frac{dT}{dH}\right)_p &=& \frac{-1}{{T^n}}\,\frac{\Delta m}{\Delta\gamma}\ .
\label{eq:5b}
\eea
In addition, we obtain from the equilibrium condition in conjunction with Eq.~(\ref{eq:2b}) a
third CC equation,
\be
\left(\frac{dH}{dp}\right)_T = \frac{\Delta v}{\Delta m}\ .
\label{eq:5c}
\ee
\ese
These three CC equations are the basis of our discussion.

Let us start by briefly discussing the obvious coexistence region in the $T$-$p$ plane, which is labeled ``{ordered}'' in Fig.~\ref{fig:1}.
The first-order transition across this plane, which is driven by the external field, does not involve any change in either entropy
or specific volume. We thus have $\Delta s = \Delta v = 0$. Equations (\ref{eq:5b}, \ref{eq:5c}) then imply $(dH/dT)_p = (dH/dp)_T = 0$.
This identifies the $H=0$ plane as the locus of the coexistence curves. $(dT/dp)_{H=0}$, Eq.~(\ref{eq:5a}), is indeterminate,
which is consistent with the fact that {\em any} curve in the $H=0$ plane below the {transition} temperature is a coexistence curve.
The CC equations thus correctly describe the coexistence plane, but do not provide any nontrivial information.

This changes as we consider the other coexistence surfaces, viz., the tricritical wings. Obviously, we have $\Delta m > 0$ 
across any first-order transition, but now $\Delta s$ and $\Delta v$ will not be zero. To find $\Delta v$, we turn to scaling theory. 
{Scaling is often thought of as valid only at second-order transitions. However, Fisher and Berker \cite{Fisher_Berker_1982} have
shown that finite-size scaling considerations allow for the definition of a diverging length scale even at a first-order transition.
Consequently, a classical first-order transition can be considered a limiting case of a second-order transition, and the homogeneity
laws, exponent relations, etc., that are known from the scaling description of second-order transitions still hold. This formalism has
recently been generalized to the case of quantum first-order transitions \cite{Kirkpatrick_Belitz_2015}, and we now apply it to the
problem under consideration.}
Let $r = (p - p^*)/p^*$ be the dimensionless distance from the transition at $T=0$. Then the generalized 
Gibbs free energy obeys a homogeneity law~\cite{Kirkpatrick_Belitz_2015}
\be
{\tilde g}(r,H,T) = b^{-(d+z)}\,\Phi_{\tilde g}(r\,b^{d+z}, H\,b^{d+z}, T\,b^z)\ .
\label{eq:6}
\ee
Here $b>0$ is the length scaling parameter, $\Phi_{\tilde g}$ is a scaling function, and we have made use of several exponent
values that characterize a first-order quantum phase transition (see Ref.~\onlinecite{Kirkpatrick_Belitz_2015}
for details): {$z$ is the relevant dynamical exponent \cite{multiple_z_footnote}}, the inverse correlation length exponent has 
its largest possible value
$1/\nu = d + z$, and the scale dimension of the field, $[H] = \beta\delta/\nu$, reflects the fact that the order-parameter exponents
are $\beta = 0$, $\delta = \infty$, such that $\beta\delta = 1$. This is a generalization of the scaling description of classic first-order
transitions given by Fisher and Berker~\cite{Fisher_Berker_1982}. 
Differentiating {${\tilde g}$ with respect to $r\propto p$, we see that the scaling part of the specific volume, $v = \partial{\tilde g}/\partial p
= \partial{\tilde g}/\partial r$, has a zero scale dimension. This implies a discontinuity of the specific volume across the transition,
and a corresponding $\delta$-function contribution to the compressibility $\kappa = -(\partial v/\partial p)/v$. This is in direct analogy
to the latent heat at a classical first-order transition and can be interpreted as a ``latent volume'' at a pressure-driven
QPT, i.e., the volume changes by a finite amount upon an infinitesimal change in pressure  {\cite{latent_heat_footnote}}. 
We also note that differentiating with respect to $r$ again we obtain $\kappa \propto \vert r \vert^{-1}$; see Ref.~\cite{Fisher_Berker_1982}
for an interpretation of this power-law divergence at a first-order transition in terms of finite-size scaling.}
The compressibility is positive in thermodynamic equilibrium, and we thus have $\Delta v > 0$ across any coexistence curve contained 
in the wings. The only remaining question is the sign of $\Delta\gamma$. For a transition from an ordered phase to a disordered one,
one would naively expect $\Delta\gamma < 0$. We will adopt this expectation for now and give a more detailed
discussion below. 

From Eq.~(\ref{eq:5a}) we conclude that $(\partial T/\partial p)_H < 0$, i.e., the slope of any coexistence curve at fixed $H$ is negative, 
and it approaches $-\infty$ as {$1/T^n$} for $T\to 0$. Similarly, Eq.~(\ref{eq:5b}) yields $(\partial T/\partial H)_p > 0$, and it approaches
$+\infty$ as $1/T^n$ for $T\to 0$. This means that the wings are necessarily perpendicular to the $T=0$ plane, and, in particular, the
coexistence curve in zero field has an infinite slope at the QPT. Finally, Eq.~(\ref{eq:5c}) implies that $(\partial H/\partial p)_T$ is
positive and finite, which implies that the wings are tilted in the direction of the {disordered} phase and are {\em not} perpendicular to 
the $p\,$-axis. 

\begin{figure}[t]
\includegraphics[width=8cm]{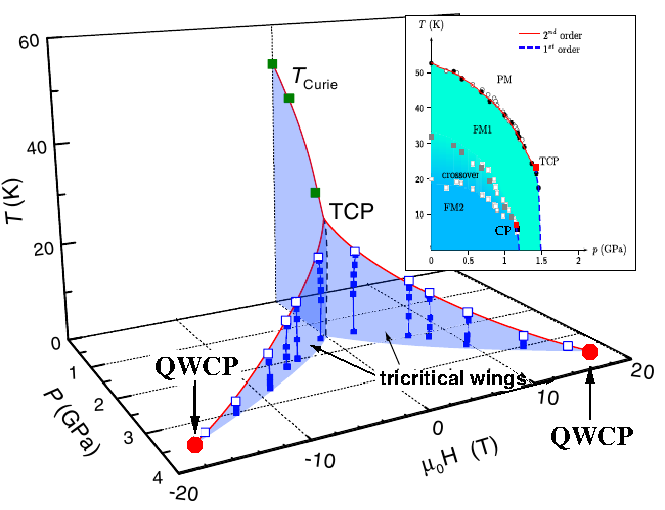}
\caption{Measured phase diagram of UGe$_2$, with the same notation as in Fig.~\ref{fig:1}. The main figure is adapted
              from Ref.~\onlinecite{Kotegawa_et_al_2011b}; the squares represent data points, the lines and surfaces are
              guides to the eye. The inset shows the $H=0$ plane with data from Ref.~\onlinecite{Taufour_et_al_2010}. Note
              the extremely steep drop of the Curie temperature past the tricritical point. The ferromagnetic phase consists of
              two phases, FM1 and FM2, separated by a line of first-order transitions at low temperatures that ends in a 
              critical point (CP).}
\label{fig:2}
\end{figure}

We now turn to an explicit example that illustrates all of the above considerations, namely, the quantum
phase transition in clean metallic ferromagnets \cite{Brando_et_al_2015, clean_footnote}. In this case, the exponent $n$
in Eqs.~\ref{eqs:5} is $n=1$, and the dynamical exponent $z$ in Eq.~(\ref{eq:6}) is $z=1$. There is a second dynamical
exponent $z=3$, but for our purposes $z=1$ yields the dominant contribution (see Refs.~\cite{Brando_et_al_2015, Kirkpatrick_Belitz_2015}
for a detailed discussion of this point). The order parameter $m$ is the magnetization, $H$ is the external magnetic field, and
the phase diagram is generically observed to have the topology shown in Fig.~\ref{fig:1}. The features discussed above
are indeed universally observed in all cases where the tricritical wings have been mapped out in detail.
As an example, we show the experimentally determined wings in UGe$_2$ in Fig.~\ref{fig:2}; for other examples, see
Ref.~\onlinecite{Brando_et_al_2015}.
Note the extremely sharp drop of the Curie temperature for pressures above the tricritical pressure that is apparent in the
inset. Also of interest is the transition inside the ferromagnetic phase (from FM1 to FM2), which is first order at
low temperatures. Our considerations apply to this transition as well, and the steep drop of the transition temperature
is again consistent with an infinite slope of the coexistence curve at $T=0$. 

We now return to the issue of the sign of $\Delta\gamma$, or more generally $\Delta s$, across the coexistence curve.
From Eq.~(\ref{eq:3}) we see that $(\partial s/\partial p)_{T,H} = -(\partial v/\partial T)_{p,H} = -v\alpha_p$, with $\alpha_p = (\partial v/\partial T)_p/v$
the thermal expansion coefficient. An increase in entropy with increasing pressure thus implies $\alpha_p < 0$. Returning to
Eq.~(\ref{eq:5a}), and remembering that $\Delta v > 0$ since the compressibility is necessarily positive, we see that a
decreasing Curie temperature with increasing pressure implies a negative thermal expansion coefficient, and vice versa.
Consistent with this, the thermal expansion coefficient at low $T$ is indeed negative in UGe$_2$ \cite{Kabeya_et_al_2010}, 
MnSi \cite{Miyake_et_al_2009}, and ZrZn$_2$ \cite{Ogawa_1983}, which all are low-$T$ ferromagnets with qualitatively
identical phase diagrams. It is interesting that $\alpha_p < 0$ by itself implies that the high-pressure phase must be
the paramagnetic one. We also note that the volume is discontinuous if the coexistence curve is crossed at fixed
$p$ as a function of $T$ as well as at fixed $T$ as a function of $p$. This is intuitively obvious and also follows from
Eqs.~(\ref{eq:3}, \ref{eq:6}). Accordingly, $\alpha_p$ at the first-order transition has
a $\delta$-function contribution that reflects the same ``latent volume'' as the corresponding $\delta$-function contribution
to the compressibility. This is consistent with the experiment by Kabeya et al~\cite{Kabeya_et_al_2010}, who observed
a pronounced negative peak in $\alpha_p$ at the transition, which they attributed to a broadened first-order transition. 

\begin{figure}[t]
\includegraphics[width=6.5cm]{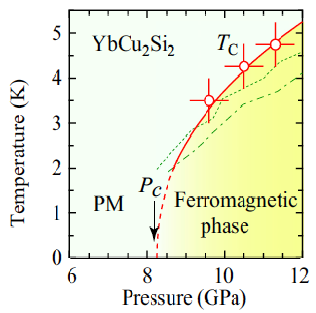}
\caption{Temperature-pressure phase diagram of YbCu$_2$Si$_2$. Ferromagnetism is induced by hydrostatic pressure
              $p>P_c \approx 8.25\,$Gpa. From Ref.~\onlinecite{Tateiwa_et_al_2014b}.}
\label{fig:3}
\end{figure}

While in most quantum ferromagnets hydrostatic pressure destroys the ferromagnetic order, there are systems in which
the opposite occurs. An example is YbCu$_2$Si$_2$, which is paramagnetic at ambient pressure, but becomes
ferromagnetic upon the application of hydrostatic pressure of roughly 10 GPa \cite{Tateiwa_et_al_2014b}, see Fig.~\ref{fig:3}. 
The thermodynamic arguments presented above then predict that $\alpha_p$ in this material, at this pressure and at low
temperature, must be positive. The thermal expansion data of Ref.~\onlinecite{Uwatoko_et_al_1993} are consistent
with this prediction, although not quite conclusive, as they focused on a higher temperature region. There also are
materials where hydrostatic pressure drives the system away from ferromagnetic order, while uniaxial stress favors it; an
example is UCoAl \cite{Shimizu_et_al_2015b}. This can be understood by realizing that in many solids the thermal
expansion coefficient is anisotropic to the point of being positive along some crystal axes, but negative along others
\cite{thermodynamics_of_solids_footnote}.

We finally briefly discuss an explicit equation of state that has been used to describe the qualitative phase diagram
of metallic quantum ferromagnets and {leads to a} schematic phase diagram as shown in 
Fig.~\ref{fig:1}~\cite{Belitz_Kirkpatrick_Rollbuehler_2005}. It is derived by minimizing a generalized Landau functional
\be
f(m) = -Hm + rm^2 + wm^4\ln(m^2 + T^2) + um^4
\label{eq:7}
\ee
with respect to the magnetization $m$. Here $m$, $H$, and $T$ are measured in suitable microscopic units,
and $r$, $u$, and $w$ are parameters of the generalized Landau theory. The physical origin of the term with
coupling constant $w$ is due to soft fermionic excitations that couple to the magnetization; this has been
discussed in detail before \cite{Belitz_Kirkpatrick_Vojta_1999, Kirkpatrick_Belitz_2012b, Brando_et_al_2015}
and will not be repeated here. In zero field, $H=0$, the logarithmic term leads to a first-order transition at
$r = r_1 = w\,e^{-1-u/w}$ where the magnetization value is $m_1 = \sqrt{r_1}$, and to a tricritical point at $T_{\text{tc}} = e^{-u/2w}$. The coexistence curve can
easily be obtained explicitly \cite{Belitz_Kirkpatrick_Rollbuehler_2005}; here we just quote the asymptotic
behavior for $r\to r_1$,
\be
T(r\to r_1) = \frac{1}{\sqrt{w}}\,(r_1 - r)^{1/2}
\label{eq:8}
\ee
which yields 
\be
dT/dr\vert_{r\to r_1} = -1/2wT\ ,
\label{eq:9}
\ee
in agreement with the general conclusions drawn above from thermodynamics, see Eq.~(\ref{eq:5a}) and
the related discussion. In the presence of a small magnetic field, one finds for the coexistence curve in
the $T=0$ plane
\be
H = m_1\left(1 + \frac{3}{11}\,\frac{u}{w}\right)\,\delta r + O(\delta r^2)\ ,
\label{eq:10}
\ee
where $\delta r = r - r_1$. This reflects the linear slope of the tricritical wings with respect to the $r$-axis that follows from
Eq.~(\ref{eq:5c}). A more involved, but elementary analysis shows that the tricritical wings are perpendicular
to the $T=0$ plane everywhere. We stress that these properties are not tied to the specific physical mechanism
that underlies the free-energy function (\ref{eq:7}); they must be true for {\em any} model that leads to a
first-order quantum phase transition and correctly reflects thermodynamics. 

We conclude with some additional discussion points. (1) The most often observed shape of the phase diagram in
quantum ferromagnets, with increasing hydrostatic pressure driving the system into the disordered phase, is not
what one might naively expect. In a fluid analogy, this is equivalent to what is observed in H$_2$O and H$_2$S,
while in most fluids increased pressure stabilizes the ordered phase. In quantum ferromagnets the latter can also
occur, see Fig.~\ref{fig:3}, but it is not common. As we have shown, this feature of the phase diagram is tied to
the sign of the thermal expansion coefficient, which tends to be negative in low-temperature ferromagnets. 
Regardless of whether hydrostatic pressure induces or destroys ferromagnetism, Eq.~(\ref{eq:5c}) implies that
the tricritical wings must always extend in the direction of the paramagnetic phase. 
(2) All actually measured tricritical wings show all of the structural features discussed above, as they must,
since the former hinge on basic thermodynamics only. However, occasionally schematic drawings of wings in the 
literature violate these thermodynamic requirements: They show wings that are perpendicular to the pressure axis 
and/or not perpendicular to the $T=0$ plane. 
(3) The discontinuity of the specific-heat coefficient across the first-order transition, $0 < \Delta\gamma < \infty$,
has an interesting implication for the dynamical critical exponents in the system. Since the static and dynamic specific-heat
exponents $\bar\alpha$ and $z_c$ are related by a hyperscaling relation ${\bar\alpha} = \nu(z_c - d)$, with $\nu$ the
correlation-length exponent and $d$ the spatial dimensionality of the system \cite{Kirkpatrick_Belitz_2015}, a
discontinuous specific-heat coefficient (${\bar\alpha} = 0$) implies $z_c = d$. This is indeed the value of $z_c$
within an explicit theory that describes the first-order transition, see Ref.~\onlinecite{Kirkpatrick_Belitz_2015}.

%[Check whether it's true that $\Delta V/V$ in MnSi (Miyake et al 2009) is much smaller than in UGe$_2$ (for the PM-FM transition, NOT the FM1-FM2 transition). If so, that might explain why $T_C$ goes to zero much more steeply in UGe$_2$ than in MnSi.]

%[Anything else? Gibbs phase rule? Can one have a triple point where, say, an FM1, and FM2, and a PM phase coexist?]

\acknowledgments
This work was supported by the NSF under grant Nos. DMR-1401410 and DMR-1401449. 

%\bibliography{Clausius_Clapeyron}

\end{document}